\documentclass{PoS}

\title{Re-evaluating the Local Compton-thick AGN Fraction }

\ShortTitle{Re-evaluating the Local Compton-thick AGN Fraction }

\author{\speaker{Angela Malizia}\\
        INAF- IASF-Bo, Via P. Gobetti 101, 40129 Bologna, Italy\\
        E-mail: \email{malizia@iasfbo.inaf.it}}
\author{J.~B. Stephen, L. Bassani,\\
          INAF-IASF Bologna, Via P. Gobetti 101, 40129 Bologna, Italy}
 \author{ A.~J. Bird,\\
 School of Physics and Astronomy, University of Southampton, SO17 1BJ, Southampton, U.K.}
 \author{F. Panessa, P. Ubertini\\
      INAF-IASF Roma, via del Fosso del Cavaliere 100, I-00133 Roma, Italy }

\abstract{We study the N$_{H}$ distribution in a complete sample of 88 AGN selected in the 20-40 keV band from 
INTEGRAL/IBIS observations.
We find that the fraction of absorbed (N$_{H} >$ 10$^{22}$ cm$^{-2}$) sources is 43\% while  Compton thick 
AGN comprise 7\% of the sample. While these estimates are fully compatible with previous soft gamma-ray surveys, 
they would appear to be in contrast with results reported from an  optically selected sample.
This apparent difference can be explained as being due to a selection bias caused by the reduction 
in high energy flux in Compton thick objects rendering them invisible at our sensitivity limit. 
Taking this into account we estimate that the fraction of highly absorbed sources is actually in close 
agreement with the optically selected sample. 
Furthermore we show that the measured fraction of absorbed sources in our sample decreases from 80\% to 
$\sim$20-30\% as a function of redshift with all Compton thick AGN having z $<$ 0.015. 
We conclude that in the low redshift bin we are seeing almost the entire AGN population, 
from unabsorbed to at least mildly Compton thick objects, while in the total sample we lose the heavily absorbed 
'counterparts' of distant and therefore dim sources with little or no absorption. 
Taking therefore this low z bin as the only one able to provide the 'true' distribution of absorption 
in type 1 and 2 AGN, we estimate the  fraction of Compton thick objects to be >24\%.}  

\FullConference{The Extreme sky: Sampling the Universe above 10 keV - extremesky2009,\\
		October 13-17, 2009\\
		Otranto (Lecce) Italy}

\begin{document}

\section{Introduction}
The cosmological evolution of the Active Galactic Nuclei (AGN) luminosity function and its implications on the
Cosmic X-ray Background (CXB) is still a challenging issue for  extragalactic science.
While years ago the study of the cosmological and statistical properties of AGN was principally limited to the 
optical  or soft X-ray regimes, and therefore dealing essentially with unabsorbed (type 1) AGN, it is now clear that 
a complete census of the entire AGN population and especially for the most obscured objects is the missing 
ingredient to come close to the true picture.
Indeed the selection and the identification of obscured objects is a difficult task in the optical as well as
in the soft X-ray (up to a few keV) band where hydrogen column densities (N$_{H}$) of the order of 
10$^{21}$-10$^{22}$ cm$^{-2}$ strongly reduce the flux emitted from the nucleus.
However, X-ray observations below 10 keV have extensively probed the so called Compton thin regime, i.e. column
densities below 1.5 $\times$ 10$^{24}$  cm$^{-2}$ (the inverse of the Thomson cross-section) but still in excess of the
Galactic value in the source direction. The Compton thick regime has been much less sampled either due
to the lack of complete spectral coverage and/or all-sky surveys above 10 keV (for mildly Compton thick
sources) or because the entire high energy spectrum is down scattered by Compton recoil and therefore
depressed at all energies (heavily Compton thick sources). Until now, indirect arguments have been used to
probe this regime: the intensity of the iron line at 6.4 keV (equivalent width typically of the order of 1 keV, Matt 1999),
the signature of strong Compton reflection, or the ratio of the observed X-ray luminosity against an isotropic
indicator of the source intensity, often the [OIII]5007 $\AA$ luminosity. However, sometimes iron line and Compton
reflection diagnostics may lead to a wrong classification, caused by a temporary switching off of the primary
continuum (Guainazzi et al. 2005) and not by thick absorption. Furthermore, the [OIII] luminosity is not
always available and/or properly estimated so that the large uncertainties on the L$_{X}$/L$_{[OIII]}$ ratios can also
lead to a misclassification. \\
The study of Compton thick AGN is important for various reasons:
 (i) about  80\% of the active galactic nuclei in the local Universe are obscured (e.g., Maiolino et al. 1998;
Risaliti et al. 1999); 
(ii) their existence is postulated in all AGN synthesis models of the X-ray background (Gilli et al 2007);
(iii) they may constitute an important ingredient for the IR and the sub-mm backgrounds, where most of
the absorbed radiation is re-emitted by dust (Fabian \& Iwasawa 1999; Brusa et al. 2001) and
(iv) accretion in these objects may contribute to the local black hole mass density (Fabian \& Iwasawa 1999, 
Marconi et al. 2004).\\
Because of this interest and despite the limitations so far encountered, a sizable sample of Compton
thick AGN is available for in depth studies (Della Ceca  et al. 2008). However, this sample is by no means complete,
properly selected and reliable in relation to the column density estimates. It is clear that for an unbiased
census of Compton thick sources sensitive soft gamma-ray surveys/observations are needed.\\

\section{The importance of hard X-ray surveys and the current scenario}
A step forward in the census of  Compton thick AGNs, is now provided by Swift/BAT and INTEGRAL/IBIS which are surveying the sky above 20 keV with a
sensitivity better than a few mCrab and a point source location accuracy of 1-3 arcmin depending on the
source strength and distance (Bird et al. 2007). These two surveys are complementary, not only because they
probe the sky in a different way but also because they can be a check of each other's results. Together they
will provide the best yet knowledge of the extragalactic sky at gamma-ray energies.
Results obtained so far from these two instruments, point to a percentage  of absorbed sources (N$_{H}>$ 10$^{22}$  
cm$^{-2}$) in the range 50-65\%, while the fraction of Compton thick objects is constrained  to be 
$<$ 20\%, likely closer to 10\% (see summary table in Ajello 2009).
This percentage is clearly in contrast with results from optically selected samples (we will, in the following, 
refer to the Risaliti et al. 1999 sample) and with that postulated in the
synthesis models of the cosmic X-ray background (Gilli et al. 2007).

\section{The INTEGRAL Complete AGN sample} 
The complete sample of INTEGRAL selected AGN has been extracted from a set of 140 extragalactic objects detected in the
20-40 keV band and listed in the 3$^{rd}$ IBIS survey (Bird et al. 2007).
Most of these objects were already identified as active galaxies in the IBIS catalogue, while others were subsequently 
classified as such thanks to follow-up  optical spectroscopy.\footnote{For optical
classification of \emph{INTEGRAL} sources, please refer to Masetti's web page at http://www.iasfbo.inaf.it/extras/IGR/main.html}\\
From this list, a complete sample has been extracted by means of the V/V$_{max}$ test (Schmidt, 1968)
i.e. assuming that the sample is distributed uniformly in space (and that
there is no evolution), it is possible to test if the sample is
complete. The test consists of comparing the volumes contained within
the distances where the sources are observed (V) with the maximum
volumes (V$_{max}$), defined as those within the distance at which
each source would be at the limit of detection. If the sample is not
complete, the expected value for $<$V/V$_{max}$$>$ is less than 0.5,
while when complete it should be equal to 0.5.
In the case of the IBIS catalogue, the sky exposure, and therefore
the limiting sensitivity is a strong function of position, as is shown in  figure 1 (left panel).
This can be taken into consideration by using the V$_e$/V$_a$ variation of the test, introduced by
Avni \& Bahcall (1980).  Once again the expected mean value
m=$<$V$_e$/V$_a$$>$ will be 0.5 when the sample is complete.\\
Figure 1 (right panel) shows the value of $<$V$_e$/V$_a$$>$ as a function of
limiting sensitivity. It can be seen that the increasing trend becomes flat above about 5.2$\sigma$ 
at which point the ratio has a value of 0.47$\pm$0.03, consistent with completeness. \\
There are 88 objects detected in the 20-40 keV band  with a significance higher than this limit and 
they form our complete sample of INTEGRAL selected AGN:  46 objects are  of type 1 (Seyfert 1-1.5, of which 
5 Narrow Line Seyfert 1s) and  33 of type 2 (Seyfert 1.8-2); only 9 Blazars  (BL Lac-QSO) are included in the catalogue. 
It is worth noting that for all the 88 objects we have class and redshift.
The 2-10 keV flux and N$_{H}$ measurements  have been collected from literature for the well studied 
objects while XRT/XMM data analysis has been performed for the new INTEGRAL AGN (IGRJ sources)
in order to get the X-ray parameters.\\ 

\begin{figure*}[th!]
\begin{center}
\includegraphics[width=5cm]{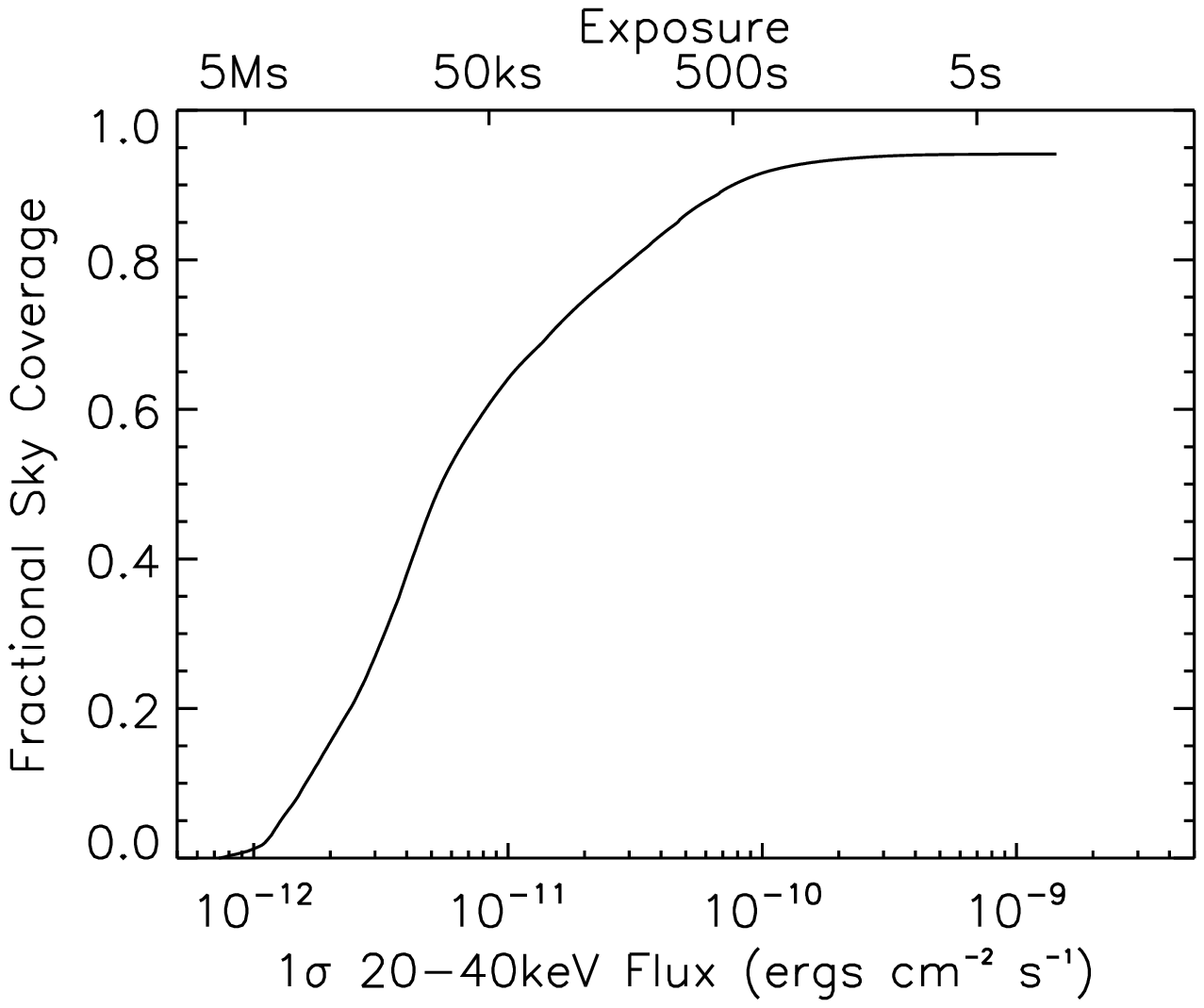}
\includegraphics[width=5cm]{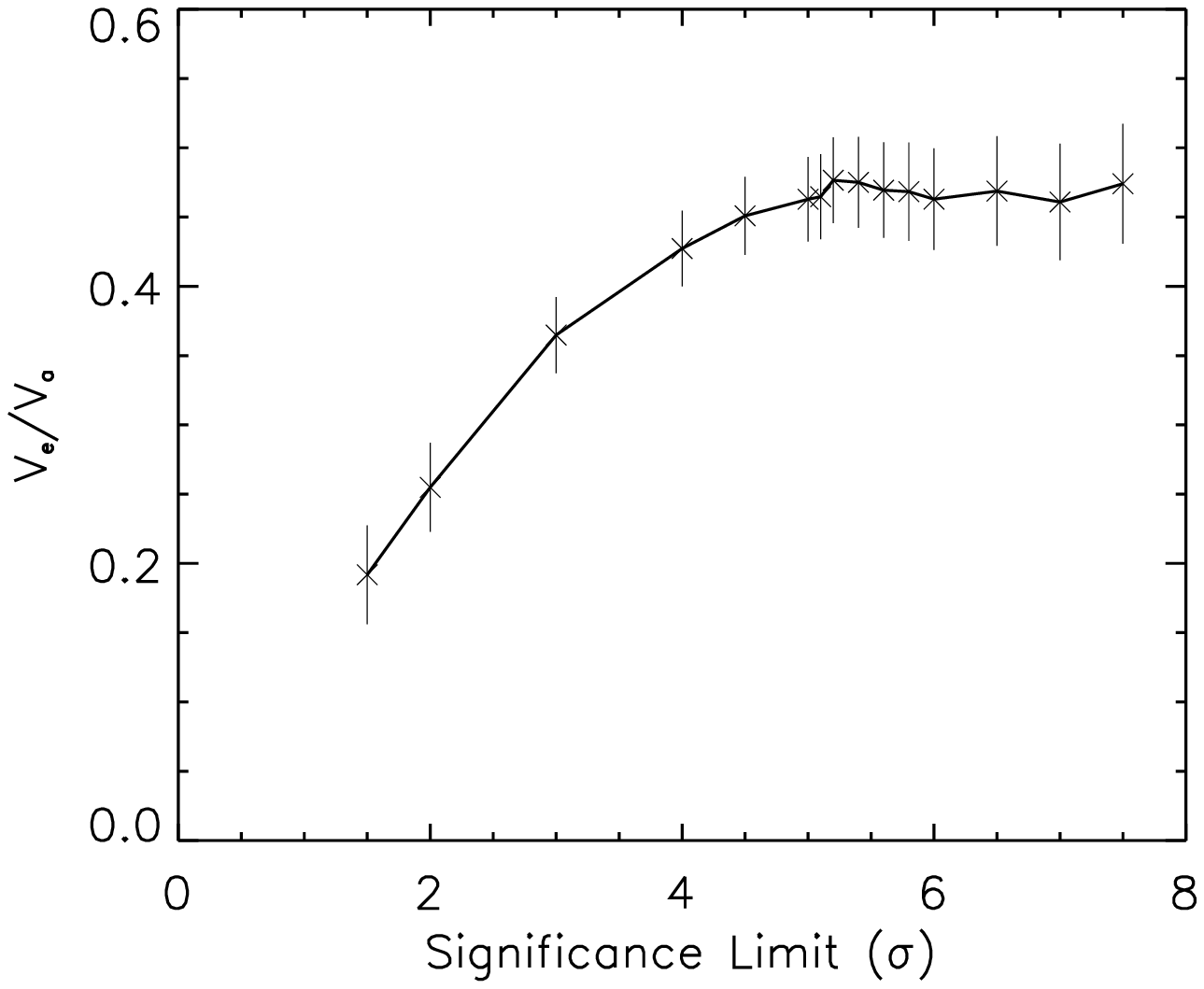}
\end{center}
\caption{\small
{{\it Left panel}:The fraction of the sky seen as a function of both 1$\sigma$ limiting flux and exposure
for the complete 3$^{rd}$ catalogue. It can be seen that large fractions of the sky 
have very different sensitivity limits.
{\it Right panel}: The value of $<$V$_e$/V$_a$$>$ as a function of limiting significance.  }}
\label{fig1}
\end{figure*}

\section{Absorption Distributions} 
The column density distribution for the complete sample is shown in figure 2 (left panel). 
Assuming N$_{H}$ = 10$^{22}$ cm$^{-2}$ as the dividing line between absorbed and unabsorbed sources, 
we find that absorption is present in 43$\%$ of the sample.  
Within our catalogue we find 5 mildly  (MKN 3, NGC 3281, NGC 4945, Circinus galaxy and  
IGR J16351-5806) and one heavily (NGC 1068) Compton thick AGN; we therefore estimate the fraction of 
Compton thick objects to be only 7$\%$. 
Although the fraction of absorbed sources is lower than obtained in various Swift/BAT and INTEGRAL/IBIS 
surveys, the percentage of Compton thick AGN is  fully consistent with these previous studies 
(see Table 1 in Ajello 2009).\\
To better investigate the absorption properties of our sample and to properly compare with optically selected 
ones, the distribution in the set of type 2 objects have also been plotted in figure 2 (right panel)
where a peak at Log N$_{H}$=23 cm$^{-2}$ is evident. Among our type 2 objects we have estimated that the fraction of 
absorbed (Log N$_{H}$ $>$22 cm$^{-2}$) is 85\% while that of Compton thick is 18\%.

\begin{figure*}[th!]
\begin{center}
\includegraphics[width=5cm]{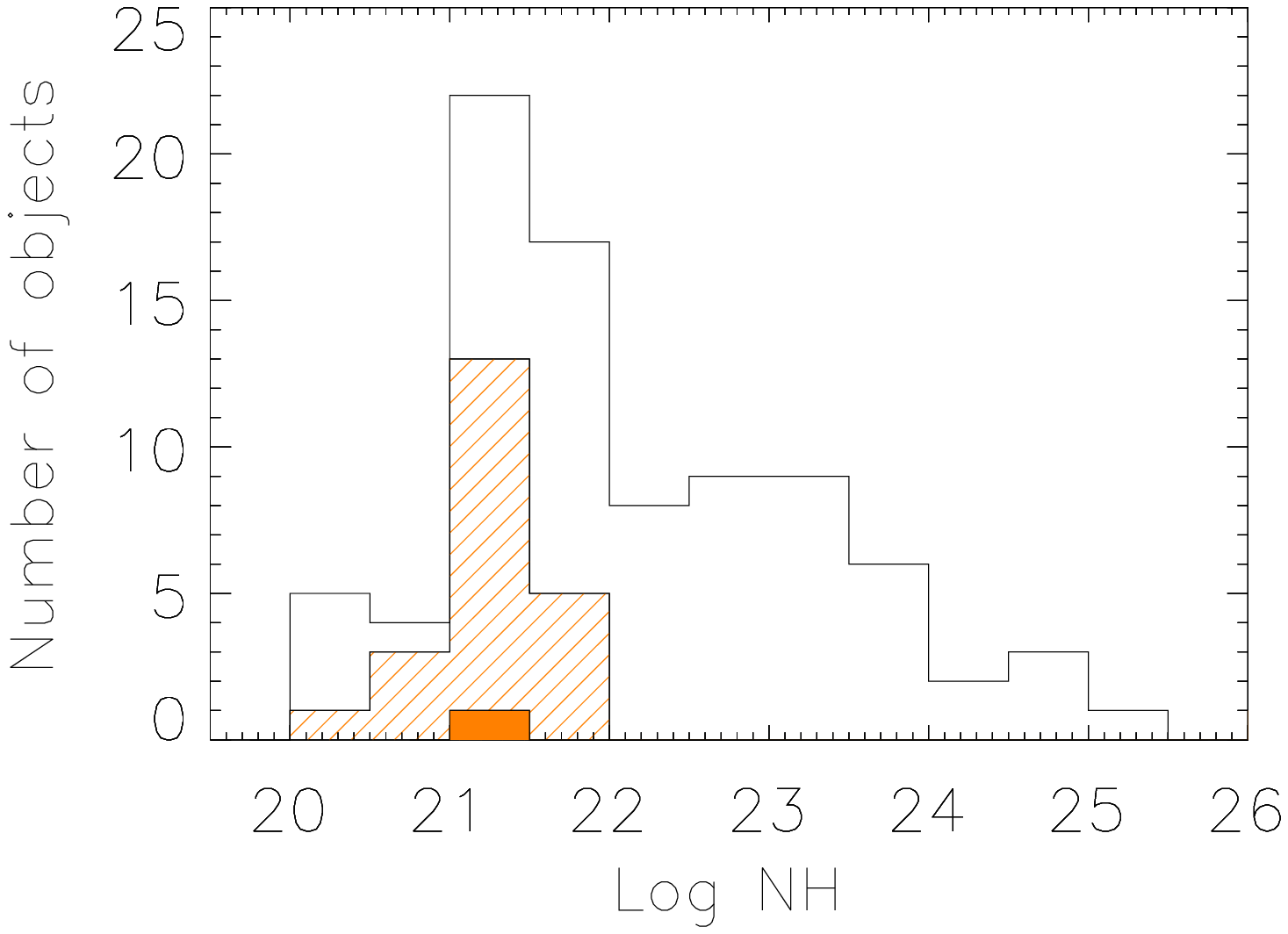}
\includegraphics[width=5cm]{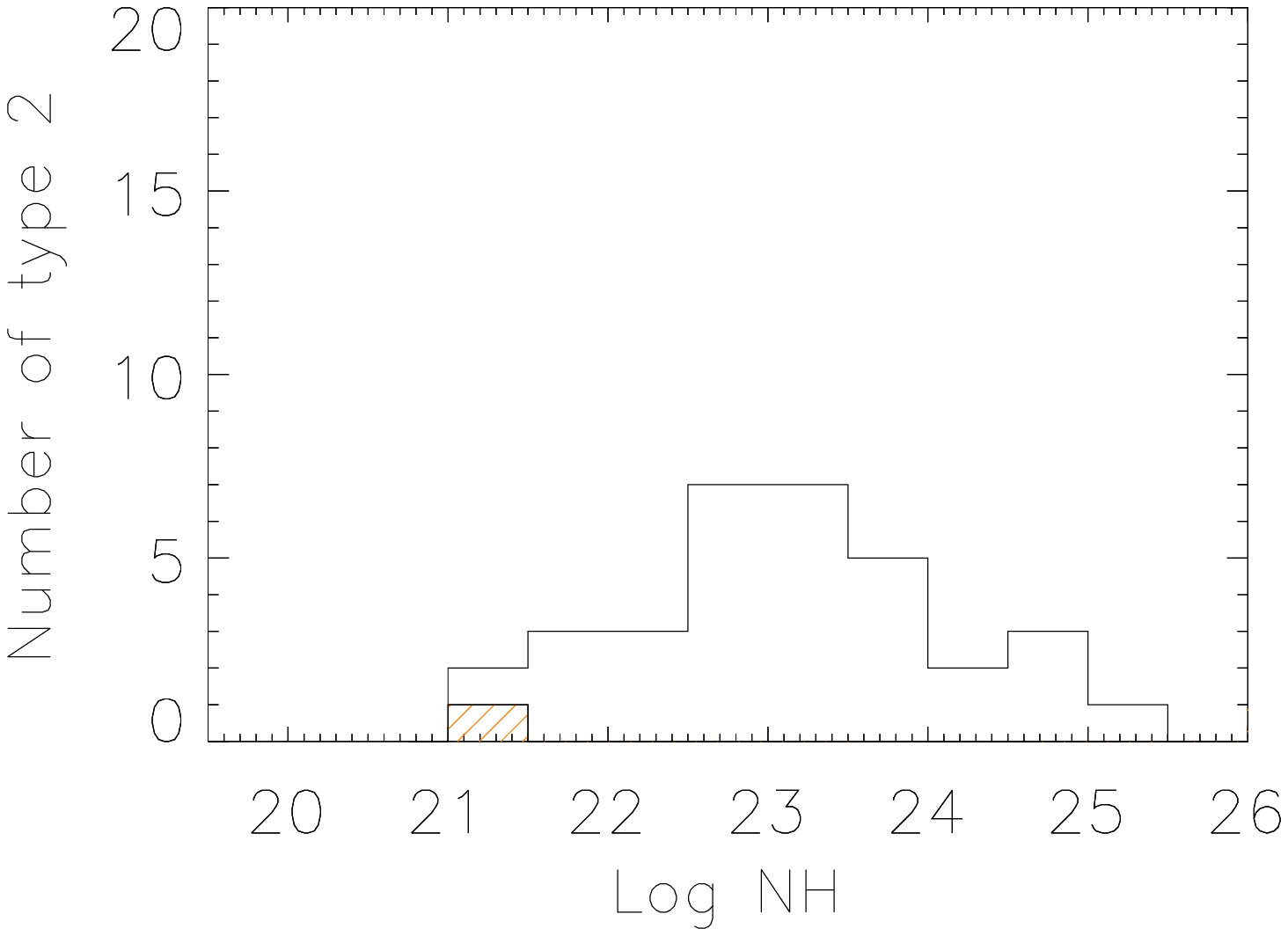}
\end{center}
\caption{\small
{{\it Left panel}:Distribution of column density in the INTEGRAL complete sample. The dashed bins represent
upper limit measurements (including Galactic values, see text), while the filled bin corresponds to GRS 1734-292 
for which a lower limit is available.
{\it Right panel}: Column density distribution in the 33 type 2 AGN of the complete sample.
Dashed bin represents IGR J16024-6107 where no absorption in excess of the Galactic one
has been measured.  }}
\label{fig1}
\end{figure*}

\section{Comparison and apparent disagreements with optically selected samples}
While the estimates of the fraction of absorbed objects as well as the fraction of Compton thick sources are
fully consistent with previous soft gamma-ray surveys, they would appear to be in contrast with results 
reported in optically selected samples.   
We will compare our results with 
the Risaliti et al. (1999) sample selected in O[III] 5007$\AA$ which is still used nowadays as a reference work in the AGN absorption 
issue. This work provided the best estimates of the key parameters of the XRB spectral intensity around the 30 keV 
peak since it relied on Beppo/SAX PDS observations of nearby bright (F$_{10-100keV}$ $>$ 10$^{-11}$ erg cm$^{-2}$  s$^{-1)}$ objects,
it also provided the first unbiased N$_{H}$ distribution of Seyfert 2 objects finding a fraction with Log(N$_{H}$)$>$22 
of 95\% and  that of Compton thick AGN of 50\%. Before comparing these results with ours, we first updated  the values of the column densities 
of the sources in the Risaliti sample, finding more recent X-ray measurements 
for many objects and for the first time an  absorption estimate for five sources.
Our re-analysis of the Risaliti sample yields a fraction of absorbed objects of 90\% but the Compton thick fraction is 36\% (15 out of 41),
i.e. smaller than found in the original paper but still a factor of two higher than our estimate.\\
It is possible that in our survey we have not recognized some Compton thick AGN because of the low statistical quality
of the X-ray observations used to estimate N$_{H}$.
To see if this has happened we can use the diagnostic diagram provided by Malizia et al. (2007).
This diagram uses  the N$_{H}$ versus  softness ratio (F$_{2-10~keV}$/F$_{20-100~keV}$) to look for AGN candidates 
and its validity has recently been confirmed by Ueda et al. (2007) and  Malizia et al. (2009).
Misclassified Compton thick objects populate the part of the diagram with low absorption and low softness ratios and none of our 
sources is located in this zone indicating that all the Compton thick sources in our complete sample have been included.\\
We have also verified that our sample, when viewed in OIII, is not significantly different to that of Risaliti et al. 
To this end, we have collected from the literature the [OIII] 5007$\AA$ fluxes for all our type 2 objects.
As noted by Maiolino and Rieke (1995) the host galaxy gaseous disk might obscure part of the narrow line region where  
the [OIII] 5007 $\AA$ emission originates. To correct for this effect we have used the prescription of Bassani et al. (1999) 
using the observed [OIII] 5007$\AA$ fluxes and Balmer decrement  H$_{\alpha}$/H$_{\beta}$ and  when the latter
was not available we based our correction on the H$_{\beta}$/H$_{\gamma}$ ratio (see Gu et al. 2006).\\
In figure 3 (left)  the distribution of [O III] 5007$\AA$ fluxes for  our sample (dashed bins) is compared with that of  Risaliti et al. (1999): 
no difference is evident from the figure indicating that we are likely  sampling the same population.\\
The most reasonable explanation for the difference in the fraction of Compton thick objects found in gamma and 
optically selected samples is due to bias introduced by obscuration which reduces the source luminosity by an 
amount depending on the column density. It is therefore more likely that, at a given distance, the most heavily 
absorbed AGN will have a flux below our sensitivity limit than unabsorbed ones and therefore will be lost from our sample.

\begin{figure*}[th!]
\begin{center}
\includegraphics[width=6.5cm]{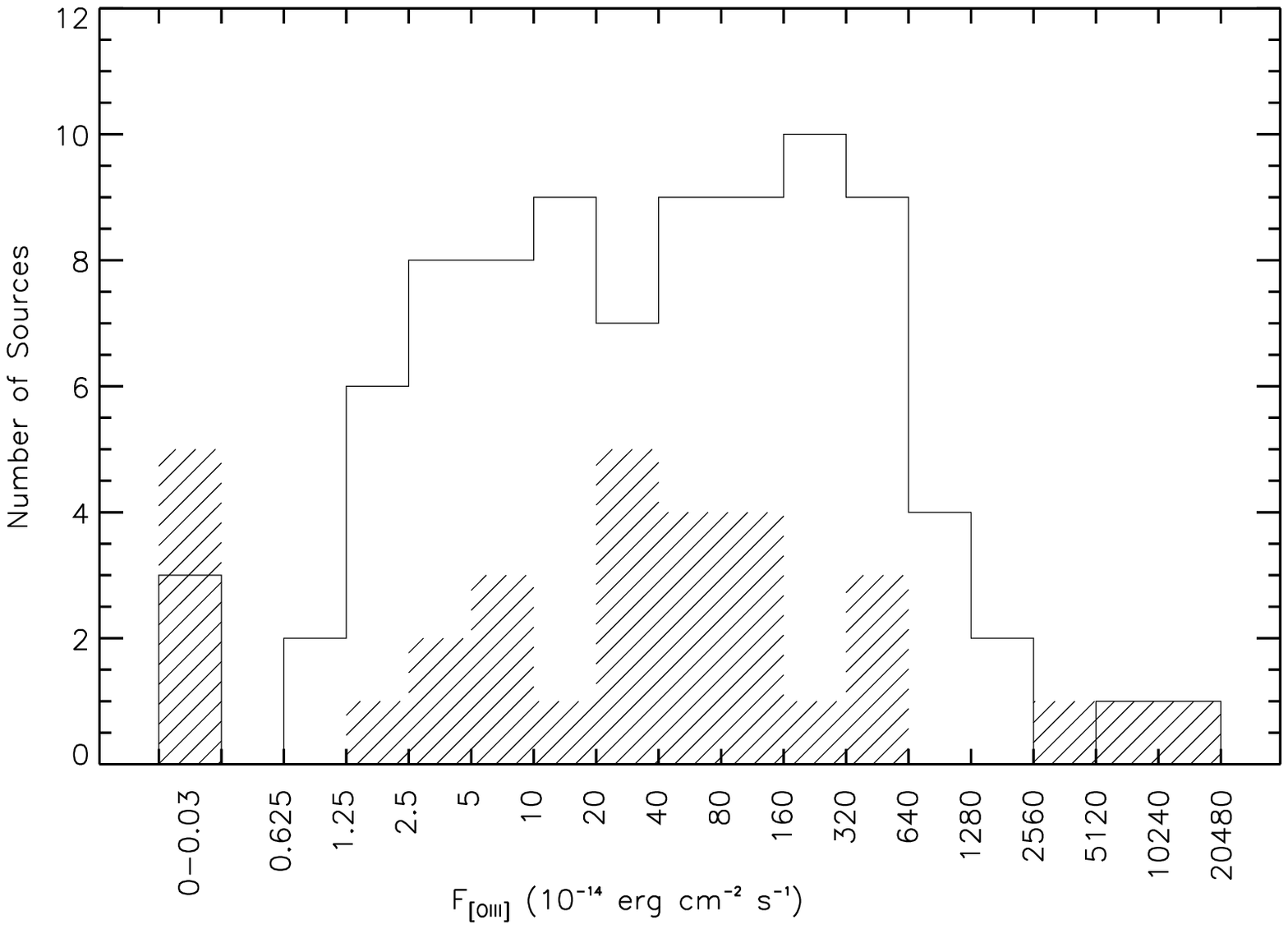}
\includegraphics[width=5cm]{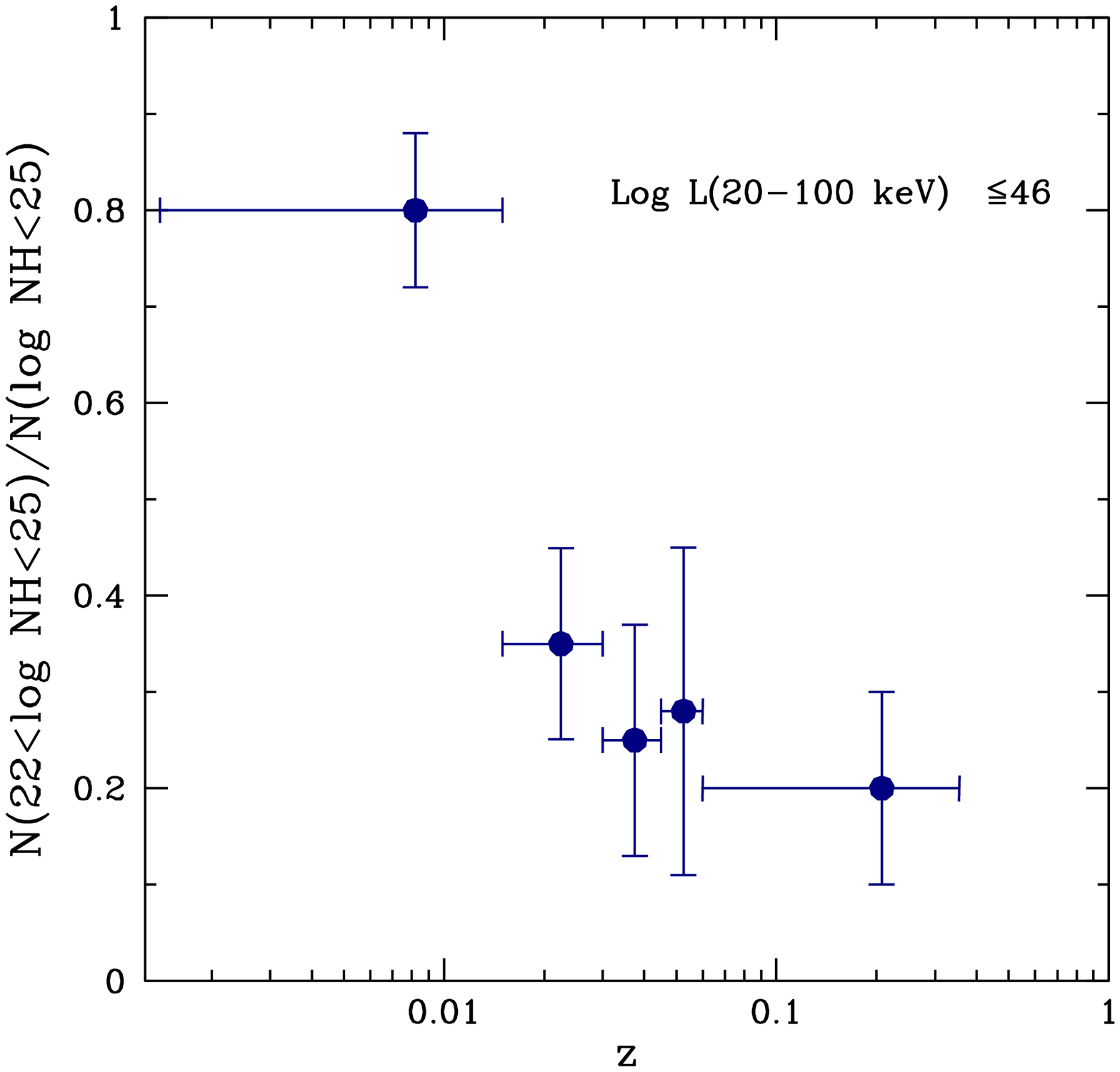}
\end{center}
\caption{\small
{{\it Left pane}: [O III] flux distribution of the Risaliti et al. sample compared to the one in INTEGRAL complete
sample (dashed bins).
{\it Right panel}: Fraction of absorbed  objects compared to the total number of AGN  as a function of redshift. }}
\label{fig1}
\end{figure*}

A method of investigating the number of these 'missing' Compton thick sources is to calculate the reduction in the 
20-40 keV flux as a function of N$_H$ using a simple absorbed power-law model in XSPEC. 
The average flux reduction is negligible below Log N$_H$=24 and becomes progressively more important thereafter 
(8\%, 25\% and 64\% reduction in the ranges 24-24.5, 24.5-25, and 25-25.5 respectively). 
Despite the simplicity of the fit adopted, the numbers do not change significantly for more complex models.
Starting from the source numbers shown in figure 2 (right panel), we can calculate that this reduction in flux would lead 
to the 'loss' of around 15 sources in the Compton thick regime assuming a Euclidian LogN/LogS. 
This suggests that the true fraction of Compton thick sources among Seyfert 2 is around 40\% in reasonable agreement with that 
found for the Risaliti et al. (1999) sample.\\
Another manner in which to examine the effect of absorption on source numbers is to calculate the fraction of absorbed 
(N$_{H}$$\geq$ 10$^{22}$ cm$^{-2}$) objects compared to the total number of AGN (i.e. the number of objects with 
N$_{H}$$\leq$ 10$^{25}$ cm$^{-2}$) as a function of redshift.
We divided our sample into 5 bins of redshift (up to z=0.335) chosen in order to have a reasonable number of sources in each bin. 
The result is shown in figure 3 (right panel)  where there is a clear trend  of decreasing fraction of absorbed objects 
as the redshift increases.
 We interpret this evidence as an indication that in the low redshift bin we are seeing almost the entire AGN population, 
from unabsorbed to at least mildly Compton thick; while in the total sample we lose the heavily absorbed 'counterparts' 
of distant and therefore dim sources with little or no absorption.\\
It is then incorrect to look at the overall sample in order to estimate the role of absorption and one manner in which 
we can come closer to the true picture is by just adopting the first redshift bin for our estimates. 
Despite the lower statistics, we are now in the position to compare our result with that of Risaliti 
et al. in a more correct way.
To do this, we use only the Seyfert 2's in our first redshift bin and then compare their column density distribution 
with that of all type 2 AGN in the  Risaliti et al. sample having z$\le$ 0.015. 
Up to this redshift,  there are 17 objects in our sample compared to 39 in that of Risaliti et al. . 
Figure 4 (left)  shows the results of this comparison: the similarity between the two distributions is 
striking  with the fraction of objects having N$_{H}$$\geq$ 10$^{23}$ cm$^{-2}$
being similar in the two samples ($\sim$ 75\%). The fraction of Compton thick objects is also remarkably close 
(35\% compared to 36\%).

\begin{figure*}[th!]
\begin{center}
\includegraphics[width=6.5cm]{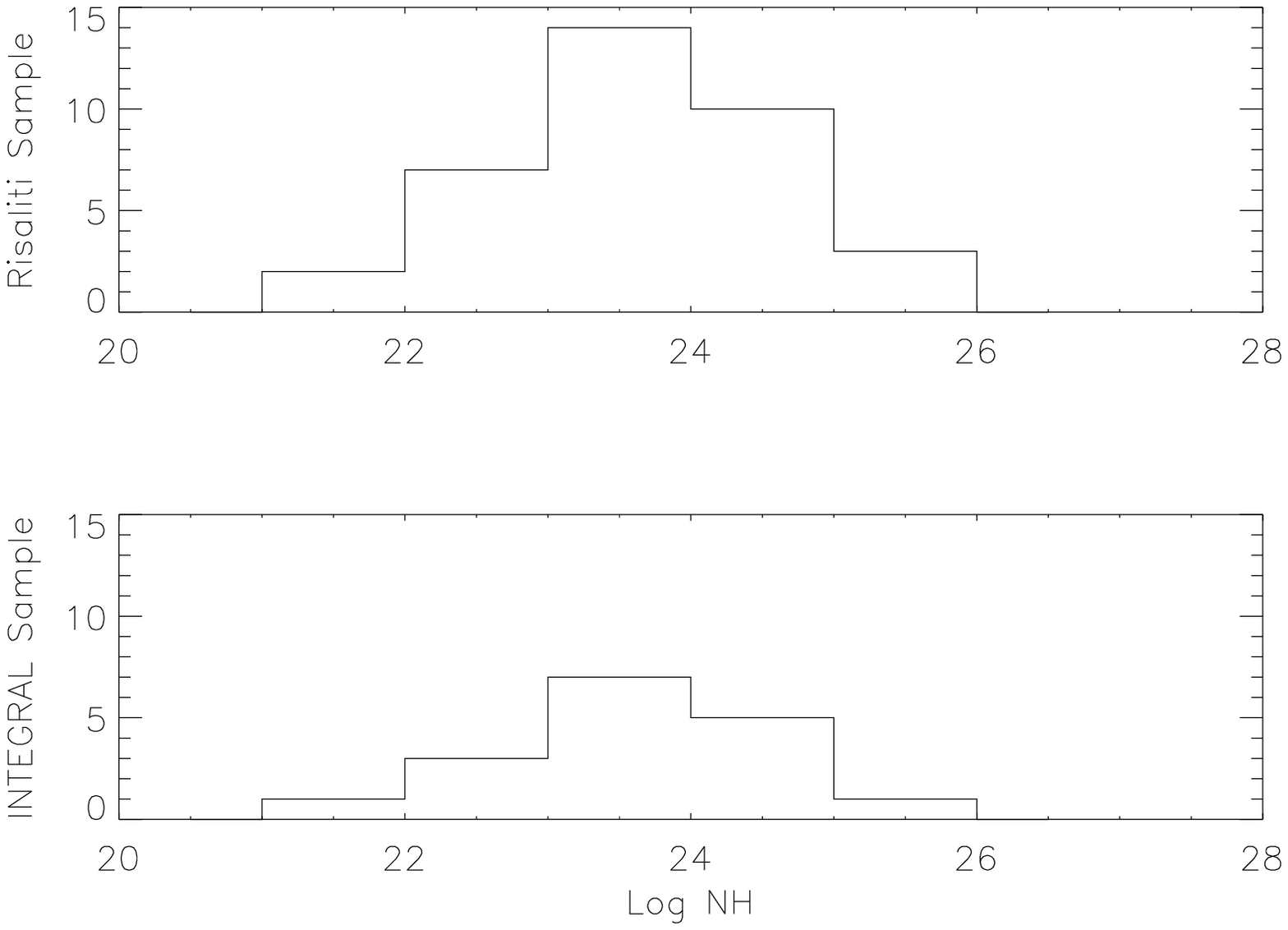}
\includegraphics[width=5cm]{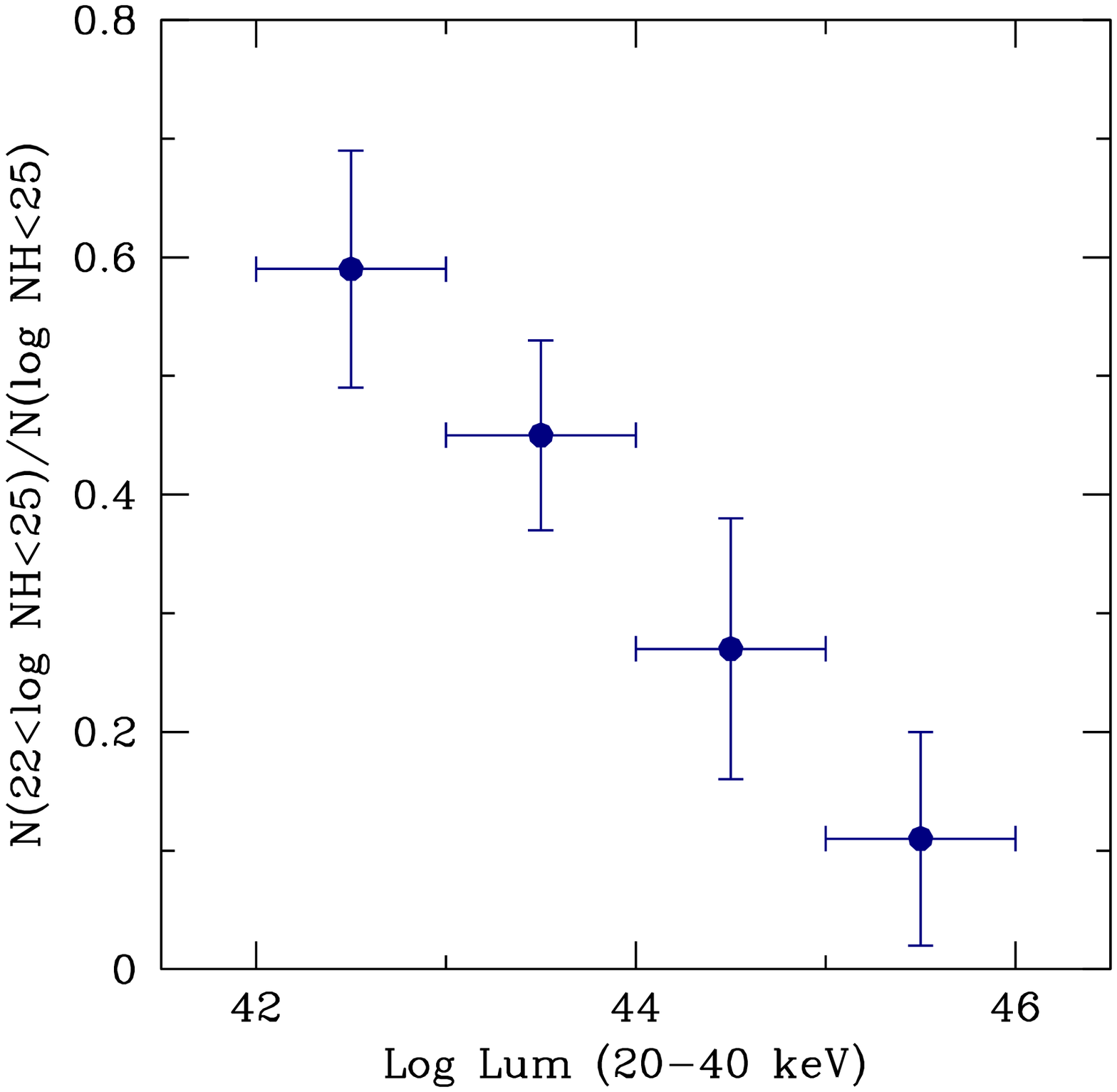}
\end{center}
\caption{\small
{{\it Left pane}: Comparison of the distribution of column densities in the type 2 objects between Risaliti et al. sample (up) and 
INTEGRAL sample (bottom) with z$\leq$0.015.
{\it Right panel}: Fraction of absorbed  objects compared to the total number of AGN  as a function of luminosity }}
\label{fig1}
\end{figure*}

In conclusion every method we use leads to an estimate of around 36\%-40\% for  the true fraction of Compton 
thick AGN among Seyfert 2.
Going from just the Seyfert 2 to the entire AGN population we note that 
the first bin, ranging up to z = 0.015, contains 25 AGN, of which 20 (80\%)  are absorbed and of these, 6 (24\%)
are Compton thick. 
It is still possible that the measured fraction of Compton 
thick objects is a lower limit, since some of the most heavily absorbed  sources may  not have  sufficient luminosities 
to be detected even at the lowest redshifts.\\
We have also looked  for a trend of decreasing  
fraction of absorbed AGN with increasing  source gamma-ray luminosities. 
This effect, which is well documented in the X-ray band (La Franca et al. 2005), has also been observed in gamma-rays 
(Bassani et al. 2006, Sazonov et al. 2007 and references therein) and is also found in our sample as shown in figure 4 (right)
Whether the redshift effect discussed here may have contaminated this result or this is a direct consequence of 
the evolution of  AGN luminosity function with z, is not possible to discriminate  here with the present data.
In fact, dividing the 25 sources with z$\le$0.015 into two luminosity bins,  we find comparable fractions of absorbed sources.
This means that either our statistics  are  too low for a proper estimate or the effect is not real but only induced by the selection due to z.
Only with the larger AGN sample that is now becoming  available from the 4th IBIS survey (Bird et al. 2009), which will be even better if  combined with
Swift-BAT extragalactic survey, will we  be able to go deeper at higher redshifts and 
provide the statistics  which will allow us to discriminate between these two effects (Malizia et al. in preparation). 
Whatever the overall picture will be it is now clear from this work that the fraction of the Compton thick objects in the local
Universe is 1 for every 4 AGN.


\begin{thebibliography}{}
\bibitem{}Ajello, M. 2009, arXiv:0902.3033
\bibitem{}Avni, Y., \& Bahcall, J. N. 1980, ApJ, 235, 694
\bibitem{}Bassani, L., Dadina, M., Maiolino, R., Salvati, M., Risaliti, G., Della Ceca, R., et al. 1999, ApJS, 121, 473
\bibitem{}Bassani, L.; Malizia, A.; Stephen, J. B., et al. proceeding of 6th INTEGRAL Workshop "The Obscured Universe" 2006, astro.ph.10455
\bibitem{}Bird, A. J., Malizia, A., Bazzano, A., et al. 2007, ApJS, 170, 175
\bibitem{}Bird, A. J., Bazzano, A., Bassani, L. et al. 2009, arXiv0910.1704
\bibitem{}Brusa, M.,  Comastri, A., Vignali, C. 2001, cghr.confE, 62B
\bibitem{}Della Ceca, R., Caccianiga, A., Severgnini,  et al. 2008, MmSAI, 79, 65
\bibitem{}Fabian, A.~C. \& Iwasawa, K. 1999, MNRAS, 303, 34
\bibitem{}Gilli, R., Comastri, A., Hasinger, G. 2007, A\&A 463, 79
\bibitem{}Gu, Q., Melnick, J., Cid Fernandes, R., Kunth, D., Terlevich, E., Terlevich, R. 2006, MNRAS, 366, 480
\bibitem{}Guainazzi, M.; Matt, G.; Perola, G.C. 2005, A\&A, 444, 119.
\bibitem{}La Franca, F., Fiore, F., Comastri, A., et al. 2005, ApJ, 635L, 864
\bibitem{}Maiolino, R., Rieke, G. H. 1995, ApJ, 454, 95
\bibitem{}Maiolino, R.; Salvati, M.; Bassani, L., Dadina, M., Della Ceca, R., Matt, G., Risaliti, G.; Zamorani, G. 1998, A\&A, 338, 781
\bibitem{}Malizia, A., Landi, R., Bassani, L., Bird, A.~J., Molina, M., et al. 2007, ApJ, 668, 81
\bibitem{}Malizia, A., Bassani, L., Panessa, F., De Rosa, A., Bird, A.~J., 2009, MNRAS, 394L, 121
\bibitem{}Marconi, A., Risaliti, G., Gilli, R., Hunt, L. K., Maiolino, R., Salvati, M. 2004, MNRAS, 351, 169 
\bibitem{}Matt, G. 1999, Nucl. Phys. B, Proc. Suppl.,  69,  467
\bibitem{}Risaliti, G., Maiolino, R.; Salvati, M. 1999, ApJ, 522, 157
\bibitem{}Sazonov, S., Revnivtsev, M., Krivonos, R., Churazov, E., Sunyaev, R. 2007, A\&A, 462, 57
\bibitem{}Schmidt, M. 1968, ApJ, 151, 393 
\bibitem{}Ueda Y. et al., 2007, ApJ, 665, 209
\end{thebibliography}
\end{document}